# A COMPREHENSIVE FRAMEWORK FOR DYNAMIC BIKE REBALANCING IN A LARGE BIKE SHARING NETWORK

L. Lin [a] and S. Peeta [b]
*[a] NEXTRANS Center,
Purdue University, USA
Email: lin954@purdue.edu*
*[b] Lyles School of Civil Engineering, Purdue University, USA
Email: peeta@purdue.edu*

**ABSTRACT**

A typical motorized passenger vehicle emits about 4.7 metric tons of carbon dioxide per year (US EPA, 2016). To decrease tailpipe emissions, reduce energy consumption and protect the environment, as of December 2016, roughly 1,000 cities worldwide have started bike sharing programs ('List of bicycle-sharing systems', 2017). Bike sharing can also help to solve the first mile and last mile problems (Lin, Yang and Chang, 2013). By providing a connection to other transportation modes, bike usage can seamlessly enable individual trips consisting of multiple transportation modes. Hence, bike sharing is becoming an important component of a modern, sustainable and efficient multi-modal transportation network.

In general, distributed bike sharing systems (BSSs) can be grouped into two types, dock-based BSS and non-dock BSS. In dock-based BSS, the bikes are rented from and returned to the docking stations. Examples of this BSS type can be found in US cities such as New York City, San Francisco, Chicago and Washington D.C. A non-dock BSS is designed to provide more freedom and flexibility to travelers in terms of access and bike usage. In contrast to dock-based BSS, riders are free to leave bikes wherever they want. Non-dock BSSs have been deployed in many cities in China by companies such as Ofo, Mobike and Bluegogo this year, and have rapidly become a popular travel mode for travelers. By September 2017, there were 15 bike-sharing programs in operation in Beijing, China that deployed over 2.3 million bikes.

While bike sharing can greatly enhance urban mobility as a sustainable transportation mode, it has key limitations due to the effects of fluctuating spatial and temporal demand. As pointed out by many previous studies (Li *et al.*, 2015; Zhou, 2015; Chen *et al.*, 2016), it is common for BSSs with fixed stations that some stations are empty with no bikes to check out while others are full precluding bikes from being returned. For non-dock BSSs, enhanced flexibility poses even more challenges to ensure bike availability at some places and prevent surplus bikes from blocking sidewalks and parking areas. To overcome these issues, this study proposes a comprehensive framework to develop optimal dynamic bike rebalancing strategies in a large bike sharing network. It consists of three models, including a *station-level pick-up/drop-off prediction model*, *station clustering model,* and *capacitated location-routing optimization model*. While the framework is proposed and illustrated for a dock-based BSS, it can easily be extended to a non-dock BSS as well.

Accurate station-level pick-up/drop-off prediction is the foundation of the proposed framework. While several prior studies have addressed pick-up and drop-off prediction, most of them focus on coarse spatial granularity, e.g., the bike sharing demand for a whole city (*Bike Sharing Demand*, 2017; Giot and Cherrier, 2014), which is not sufficient for bike rebalancing among stations. A few studies develop linear regression models for station-level demand prediction (Rixey, 2013; Faghih-Imani *et al.*, 2014). However, none of them consider underlying correlations among stations to enhance the prediction models. For example, if a bike station near a subway exit has high demand during the peak period, another one close to it may also have high demand during that period. To address these two issues, we propose a powerful deep learning model called graph convolution neural network model (GCNN) with data-driven graph filter (DDGF) (Lin *et al.*, 2017), which is especially appropriate for predictions and classifications based on data embedded in a graph network. Further, the model can automatically learn the hidden spatial-temporal correlations among stations to provide more accurate predictions. Based on the predicted pick-up/drop-off, we can calculate station-level demand gap as follows:

$$g_i^{(t+1)} = p_i^{(t+1)} - d_i^{(t+1)} - c_i^t \quad (1)$$

where, $g_i^{(t+1)}$ is the demand gap of station $i$ at time $t+1$;

$p_i^{(t+1)}$ is the pick-up prediction of station $i$ at time $t+1$;

$d_i^{(t+1)}$ is the drop-off prediction of station $i$ at time $t+1$;

$c_i^t$ is the current bike number at station $i$ at time $t$.

Station-level demand gap can either be positive or negative. We label a station with a positive gap as a bike-shortage station and that with a negative gap as a bike-surplus station.

Station clustering model is the second necessary component of the proposed comprehensive framework. It is required for a large-scale BSS with hundreds of stations. A previous study (Raviv, Tzur and Forma, 2013) showed that their model can provide optimal bike rebalancing solutions for only moderate-sized BSSs (up to 60 stations). To solve this problem, Liu et al. (2016) proposed an Adaptive Capacity Constrained K-centers Clustering (AdaCCKC) algorithm which splits the large bike-sharing network into small clusters before conducting inner-cluster routing optimization. Their algorithm is developed on the assumption that the absolute total demand gap of a cluster is less than the capacity of a shipping truck to guarantee the existence of feasible solutions. By contrast, in this study, we aim to build "self-balancing" clusters, which implies that the total demand gap of a cluster should be close to zero. Furthermore, the stations within the same cluster should also be spatially close. Let $G(N,E)$ be an undirected graph, where $N = \{1,...,n\}$ is the set of nodes and $E$ is the set of edges. It is built based on station-level demand gaps and the spatial distances between stations, as follows:

$$W_{ij}^{t+1} = \begin{cases} \dfrac{exp(-|g_i^{t+1} + g_j^{t+1}|)}{dis_{ij}}, & \text{if } g_i^{t+1} \text{ and } g_j^{t+1} \text{ have opposite signs} \\ 0, & \text{otherwise} \end{cases} \quad (2)$$

where, $W_{ij}^{t+1}$ is the edge weight between stations $i$ and $j$ at time $t+1$; $dis_{ij}$ is the spatial distance between stations $i$ and $j$.

Equation (2) illustrates that if two stations have a small net demand gap (that is, their requirements for loading and unloading bicycles are complementary) and the geographical distance between them is small, a higher edge weight will be assigned. A graph clustering algorithm labeled the Community Detection algorithm will then be applied to group stations with stronger weights together (Lin, Wang and Sadek, 2014). This clustering algorithm does not require the *a priori* determination of cluster numbers, and has been shown to be very efficient in previous studies (Blondel *et al.*, 2008).

The third model of the comprehensive framework is to solve a capacitated location-routing problem (CLRP). Most previous studies assume that the locations of bike distribution centers are known, and only focus on the shipping truck routing optimization (Forma, Raviv and Tzur, 2015). By contrast, the CLRP deals with the combination of two types of decision variables: the locations of bike distribution centers and the design of distribution routes (Lopes, Ferreira and Santos, 2016). Given a self-balancing cluster, bike-surplus stations will be treated as potential distribution centers and each bike-shortage station will be assigned to a single distribution center. The demands of bike-shortage stations are satisfied using trucks with fixed capacity. The total cost incurred in the model includes: (i) a fixed cost when a distribution center has bike-shortage stations assigned and must be open, (ii) a fixed cost for each shipping truck used, and (iii) a travel cost for each route. The goal of the CLRP is to determine the set of distribution centers and routes that minimize the total costs.

The comprehensive framework will be evaluated based on the Citi BSS in New York City, which is a real-world large-scale bike sharing network with 272 stations. Over 28 million transactions from 2013 to 2016 have been downloaded to analyze the performance of the proposed framework.